\documentstyle[preprint,aps,epsf,floats]{revtex}

\tighten

\def\OMIT#1{{}}
\def\vslash{v\hskip-0.50em /}

\begin{document}

\preprint{\vbox{
\hbox{NT@UW-01-023}
}}

\title{Heavy-Meson Observables at One-Loop in Partially Quenched
  Chiral Perturbation Theory}
\author{ Martin J. Savage\footnote{{\tt 
savage@phys.washington.edu}}}
\address{
Department of Physics, University of Washington, 
Seattle, WA 98195-1560.}

\maketitle

\begin{abstract} 
I present one-loop level calculations of the Isgur-Wise functions
for $B\rightarrow D^{(*)}e\nu$,
of the matrix elements of isovector twist-2 operators in
$B$ and $D$ mesons, 
and the matrix elements for the radiative decays $D^*\rightarrow D\gamma$
in partially quenched heavy quark chiral perturbation theory.
Such expressions are required in order to extrapolate 
from the light quark masses used in lattice simulations 
of the foreseeable future to those of nature.
\end{abstract}

\bigskip
\vskip 8.0cm
\leftline{September 2001}

\vfill\eject

\section{Introduction}

In order to extract fundamental quantities from systems containing
heavy quarks, such as the weak mixing angles $V_{bc}$ and $V_{bu}$,
the strong interaction contributions to the observable of interest
must be determined.
This is because the standard model of electroweak interactions
is constructed in terms of quarks, gluons and leptons,
while we are only able to study the quarks via the
properties of the hadrons within which they are confined.
Significant progress was made in eliminating the strong interaction
uncertainties in some matrix elements in hadrons containing one or more heavy
quarks in the late eighties with the
discovery of the heavy quark symmetries~\cite{IW}, and the 
formulation of heavy quark effective theory~\cite{HQET}.
Ultimately, such 
strong interaction effects will be directly computable
from numerical lattice QCD simulations, however,
at this point in time, and in the foreseeable future, lattice
simulations will not be performed with the physical values of the
light-quark masses, $m_q$.
In order to make a connection between the lattice simulations and
nature, an extrapolation from the lattice quark masses to those of
nature is required.
One can most efficiently perform the extrapolation for many
observables using simulations 
of partially quenched QCD (PQQCD) matched to partially
quenched chiral perturbation theory 
(PQ$\chi$PT)~\cite{BG93,SZ96,S97,GL98,SS00a,SS00b,SS01a}.
Some observables have been analyzed with PQ$\chi$PT in the light
pseudo-scalar meson sector, such as decay constants and
masses~\cite{S97}.
Further, the 
decay constants and B-parameters for mesons containing a single heavy
quark have been computed~\cite{SZ96}.
In this work we present the one-loop analysis in PQ$\chi$PT
of the Isgur-Wise function
for $B\rightarrow D^{(*)}$ decays, of the matrix elements of isovector 
twist-2 operators that are directly related to moments of the parton
distribution functions, and of the radiative decays
$D^*\rightarrow D\gamma$.

\section{PQQCD}

We will consider a theory constructed from three valence-quarks,
$u,d,s$,
three-sea quarks $j,l,r$ and three bosonic-quarks $\tilde u, \tilde
d, \tilde s$.  The masses of the bosonic-quarks are equal to those of
the  valence-quarks, $\overline{m}, \overline{m}, m_s$ respectively, where
we have chosen to work in the isospin limit.
The sea-quarks are taken to be degenerate with mass $m_j$.
The valence-, sea-, and bosonic-quarks are combined into
column vectors
\begin{eqnarray}
Q_L & = & \left(u,d,s,j,l,r,\tilde u,\tilde d,\tilde s\right)^T_L
\ \ ,\ \ 
Q_R \ =\  \left(u,d,s,j,l,r,\tilde u,\tilde d,\tilde s\right)^T_R
\ \ \ ,
\label{eq:quarkvec}
\end{eqnarray}
which transform in the fundamental representation of $SU(6|3)_{L,R}$
respectively.
The ground floor of $Q_L$ transforms as a $({\bf 6},{\bf 1})$ of
$SU(6)_{qL}\otimes SU(3)_{\tilde q L}$ while the first floor transforms
as $({\bf 1},{\bf 3})$.
The right handed field $Q_R$ transforms analogously.
The PQQCD Lagrange density is invariant under $SU(6|3)_L\otimes
SU(6|3)_R$ chiral transformations, and in direct analogy with QCD a
chiral Lagrangian can be constructed to describe the low-energy
dynamics of the low lying hadrons.

The low-energy QCD dynamics of hadrons containing a single heavy quark 
are described by a chiral lagrangian that has the heavy quark spin 
symmetry and
flavor symmetry manifest~\cite{HMchirala,HMchiralb}.
The extension of the heavy quark chiral Lagrange density to describe 
heavy quark systems in PQQCD can be found in Ref.~\cite{SZ96}.
The dynamics of the pseudo-scalar mesons are described  at leading order
by a Lagrange density of the form,
\begin{eqnarray}
{\cal L } & = & 
{f^2\over 8} {\rm
  str}\left[\ \partial^\mu\Sigma^\dagger\partial_\mu\Sigma\ \right]
 \ +\ \lambda\ {\rm str}\left[\ m_Q\Sigma^\dagger + m_Q\Sigma\ \right]
\ +\ 
\alpha_\Phi\partial^\mu\Phi_0\partial_\mu\Phi_0\ -\ m_0^2\Phi_0^2
\ \ \ \ ,
\label{eq:lagpi}
\end{eqnarray}
where the meson field is incorporated in $\Sigma$ via
\begin{eqnarray}
\Sigma & = & \exp\left({2\ i\ \Phi\over f}\right)
\ =\ \xi^2
\ \ \ ,\ \ \ 
\Phi \ =\  \left(\matrix{ M &\chi^\dagger \cr \chi &\tilde{M} }\right)
\ \ \ ,
\label{eq:phidef}
\end{eqnarray}
where $M$ and $\tilde M$ are matrices containing bosonic mesons while
$\chi$ and $\chi^\dagger$ are matrices containing fermionic mesons,
with
\begin{eqnarray}
M & = & 
\left(\matrix{
\eta_u & \pi^+ & K^+ & J^0 & L^+ & R^+\cr
\pi^- & \eta_d & K^0 & J^- & L^0 & R^0\cr
K^- & \overline{K}^0 & \eta_s & J_s^- & L_s^0 & R_s^0\cr
\overline{J}^0 & J^+ & J_s^+ & \eta_j & Y_{jl}^+ & Y_{jr}^+\cr
L^- & \overline{L}^0 & \overline{L_s}^0 & Y_{jl}^- & \eta_l & Y_{lr}^0\cr
R^- & \overline{R}^0 & \overline{R_s}^0 & Y_{jr}^- & \overline{Y}_{lr}^0 & \eta_r }
\right)
\ \ \ ,\ \ \ 
\tilde M \ =\  \left(\matrix{\tilde\eta_u & \tilde\pi^+ & \tilde K^+\cr
\tilde\pi^- & \tilde\eta_d & \tilde K^0\cr
\tilde K^- & \tilde{\overline{K}^0} & \tilde\eta_s
}\right)
\nonumber\\
\chi & = & 
\left(\matrix{\chi_{\eta_u} & \chi_{\pi^+} & \chi_{K^+} & 
\chi_{J^0} & \chi_{L^+} & \chi_{R^+}\cr
\chi_{\pi^-} & \chi_{\eta_d} & \chi_{K^0} &
\chi_{J^-} & \chi_{L^0} & \chi_{R^0}\cr
\chi_{K^-} & \chi_{\overline{K}^0} & \chi_{\eta_s} & 
\chi_{J_s^-} & \chi_{L_s^0} & \chi_{R_s^0} }
\right)
\ \ \ \ .
\label{eq:mesdef}
\end{eqnarray}
where the upper $3\times 3$ block of $M$ is the usual octet of
pseudo-scalar mesons while the remaining entries correspond to mesons
formed with the sea-quarks.
The singlet field is defined to be 
$\Phi_0 ={\rm str}\left(\ \Phi\ \right)/\sqrt{2}$,
and its mass $m_0$ can be taken to
be of order the scale of chiral symmetry breaking, 
$m_0\rightarrow\Lambda_\chi$~\cite{SS01a}.
The super mass matrix, $m_Q$, is
\begin{eqnarray}
m_Q& = & {\rm diag}\left(\overline{m},\overline{m},m_s,m_j,m_j,m_j,
\overline{m},\overline{m},m_s\right)
\ \ \ ,
\label{eq:massmat}
\end{eqnarray}
where we will work in the limit of exact isospin symmetry.
The convention we use corresponds to $f\sim 132~{\rm MeV}$.

The B-mesons with quantum numbers of $b\overline{Q}$
form a nine-component vector, 
\begin{eqnarray}
B & = & \left( B_u,B_d,B_s,B_j,B_l,B_r,B_{\tilde u},B_{\tilde
    d},B_{\tilde s}\right)
\ \ \ ,
\label{eq:Bdef}
\end{eqnarray}
and heavy quark spin symmetry (for a comprehensive review see
Ref.~\cite{ManWise})
is implemented by combining the annihilation operators for the 
$B$ and $B^*$ mesons together into the field $H_v$
\begin{eqnarray}
H_v& = & {1\over 2}(1+\vslash) \left[\ \gamma^\alpha B^*_\alpha 
+ i \gamma_5 B\ \right]
\ \ \ ,
\end{eqnarray}
where $H_v\rightarrow H_v V^\dagger$ under $SU(3)$ flavor transformations
and $H_v\rightarrow S_b H_v$ under heavy quark spin transformations.
The four-vector $v^\mu$ is the four-velocity of the heavy meson.
The low-momentum strong interaction
dynamics of the heavy mesons are described by a Lagrange
density of the form~\cite{SZ96},
\begin{eqnarray}
{\cal L} & = & 
-i {\rm Tr}\left[\ \overline{H}_v v^\mu \left(\partial_\mu H_v  
+ i H_v V_\mu\right)\ \right]
\ -\ g_\pi {\rm Tr}\left[\ \overline{H}_v H_v \gamma^\nu\gamma_5 A_\nu
\ \right]
\ \ \ ,
\label{eq:lagH}
\end{eqnarray}
where 
\begin{eqnarray}
\overline{H}_v & = & \gamma^0 H_v^\dagger\gamma^0\ =\ 
\left[\ \gamma^\alpha B_\alpha^{*\dagger}
\ +\ i\gamma_5 B^\dagger\ \right]
{1\over 2}(1+\vslash)
\ \ \ ,
\end{eqnarray}
and where the light-meson fields are
\begin{eqnarray}
A_\mu & = & {1\over 2}\left(\xi\partial_\mu\xi^\dagger
 - \xi^\dagger\partial_\mu\xi\right)
\ \ ,\ \ 
V_\mu \ =\  {1\over 2}\left(\xi\partial_\mu\xi^\dagger
 + \xi^\dagger\partial_\mu\xi\right)
\ \ \ .
\end{eqnarray}
The ${\rm Tr}\left[\ \right]$'s in eq.~(\ref{eq:lagH}) correspond to
traces over Dirac indices, and  implicit in eq.~(\ref{eq:lagH})
are ${\rm str}\left[\ \right]$'s over the flavor indices.
Couplings to $\Phi_0$~\cite{SZ96}, such as 
${\rm str}\left[\ A_\mu\ \right]$, have not been included 
as the $\Phi_0$ will be integrated out of the theory, and such
contributions will be included via
higher dimension local operators.
The axial coupling constant $ g_\pi$ has been constrained to be 
$g_\pi\sim 0.56$~\cite{radmany} (or $g_\pi = 0.24$) 
by the radiative decays 
$D^*\rightarrow D\gamma$~\footnote{A more detailed study~\cite{Stewart98} gives
$g_\pi \sim 0.76\pm 0.2$ or $g_\pi \sim 0.27\pm 0.07$.},
and recent quenched lattice simulations yield
$g_\pi\sim 0.42$~\cite{glatta,glattb}.
The low-momentum strong interaction dynamics of 
$D^{(*)}$ mesons are described in an analogous way, with heavy
quark flavor symmetry dictating that the axial coupling, $g_\pi$,
for the $B$'s and $D$'s is the same.
Further, the dynamics of the anti-heavy mesons can be described in
analogous way~\cite{SZ96}.


\section{Decay Constants}
The heavy meson decay constants, such as $f_{B_u}$, have been analyzed
at the one-loop level in PQ$\chi$PT by Sharpe and Zhang~\cite{SZ96}.
Their analysis provides a very clear demonstration of the
implementation of PQQCD and PQ$\chi$PT, and it is worth 
reviewing their results.

The decay constants of the heavy mesons are defined via the matrix element
\begin{eqnarray}
\langle 0|\overline{q}\gamma^\mu (1-\gamma_5)b|B_q (v)\rangle
 & = & 
-i \ f_{B_q}\  m_B\  v^\mu
\ \ \ .
\label{eq:decaymat}
\end{eqnarray}
In the heavy meson effective field theory, this matrix element
receives contributions from both tree-level operators and from
diagrams involving light-meson loops.
Up to ${\cal O}\left(m_q\right)$, the local operators that contribute are
\begin{eqnarray}
L_\mu^a  & = & - {\kappa\over 2} {\rm Tr}\left[\ 
\gamma_\mu (1-\gamma_5)H_v \xi^\dagger n^a\right]
\ -\ {C_1(\Lambda)\over 2} {\rm Tr}\left[\ 
\gamma_\mu (1-\gamma_5)H_v {\cal M}_+ \xi^\dagger n^a\right]
\nonumber\\
& & 
\ -\ {C_2(\Lambda)\over 2} {\rm Tr}\left[\ 
\gamma_\mu (1-\gamma_5)H_v \xi^\dagger {\cal M}_+ n^a\right]
\ -\ {C_3(\Lambda)\over 2}
{\rm Tr}\left[\ 
\gamma_\mu (1-\gamma_5)H_v \xi^\dagger n^a\right]
{\rm str}\left[\ {\cal M}_+\ \right]
\ \ \ ,
\end{eqnarray}
where $n^a$ is a column vector that picks out the light quark
associated with the desired current,  $\Lambda$ is the
renormalization scale,
and ${\cal M}_+$ is related to the mass matrix via
${\cal M}_\pm~=~{1\over 2}\left(\xi m_Q \xi \pm 
\xi^\dagger m_Q\xi^\dagger\right)$.
At leading order one finds that $f_{B_q} = \kappa/\sqrt{m_{B_q}}$.
At ${\cal O}\left(m_q\right)$ in PQ$\chi$PT
the decay constants are~\cite{SZ96}
\begin{eqnarray}
\sqrt{m_{B_q}}\ f_{B_q}^{PQ} & = &
\kappa \left[\ 1 \ -\ 
{1+3 g_\pi^2\over 32\pi^2 f^2}
\left(\ 3 m_{jq}^2\log {m_{jq}^2\over\Lambda^2} - 
{1\over 3}(2 m_{qq}^2-m_{jj}^2)\log {m_{qq}^2\over\Lambda^2}\right)
\right]
\nonumber\\
& & + \left(\ C_1^{(R)}(\Lambda)+C_2^{(R)}(\Lambda)\ \right)\ m_q 
\ +\  3 C_3^{(R)}(\Lambda)\  m_j
\ \ \ ,
\label{eq:PQfB}
\end{eqnarray}
where $m_{qq}$, $m_{jj}$ and $m_{jq}$
are the masses of  light mesons with quark composition
$\overline{q}q$, $\overline{j}j$ and $\overline{j}q$ respectively.
The $C_i^{(R)}(\Lambda)$ are renormalized constants where
both divergences and terms that are analytic in $m_q$ have been 
absorbed.
The expression in eq.~(\ref{eq:PQfB}) holds for all $q=u,d,s,j,l,r$,
and agrees with the expression in Ref.~\cite{SZ96} when there are three
sea quarks, $N_f=3$.
Lattice simulations of $f_{B_{u,d,s}}$ and $f_{B_j}$ at various values of
the sea-quark masses, $m_j$, and various values of the valence masses 
will allow for the determination of the quantities
$\kappa$, $C_3^{(R)}(\Lambda)$ and the combination
$C_1^{(R)}(\Lambda)+C_2^{(R)}(\Lambda)$.
As these counterterms have the same value in QCD and PQQCD, 
their values can be inserted into the 
QCD expressions for the decay constants~\cite{SZ96,GJMSW92}
\begin{eqnarray}
\sqrt{m_{B_u}}\ f_{B_u} & = &
\kappa \left[\ 1 \ -\ 
{1+3 g_\pi^2\over 32\pi^2 f^2}
\left(\ {3\over 2}m_\pi^2\log {m_\pi^2\over\Lambda^2}
+m_K^2\log{m_K^2\over\Lambda^2}
+{1\over 6}m_\eta^2\log{m_\eta^2\over\Lambda^2}
\right)
\right]
\nonumber\\
& & + \left(\ C_1^{(R)}(\Lambda)+C_2^{(R)}(\Lambda)\ \right)\ \overline{m}
\ +\  C_3^{(R)}(\Lambda)\ (2\overline{m}+m_s)
\nonumber\\
\sqrt{m_{B_s}}\ f_{B_s} & = &
\kappa \left[\ 1 \ -\ 
{1+3 g_\pi^2\over 32\pi^2 f^2}
\left(\ 2 m_K^2\log{m_K^2\over\Lambda^2}
+{2\over 3}m_\eta^2\log{m_\eta^2\over\Lambda^2}
\right)
\right]
\nonumber\\
& & + \left(\ C_1^{(R)}(\Lambda)+C_2^{(R)}(\Lambda)\ \right)\ m_s
\ +\  C_3^{(R)}(\Lambda)\ (2\overline{m}+m_s)
\ \ \ .
\label{eq:QCDfB}
\end{eqnarray}

\section{Chiral $1/m_c^2$ Corrections to $B\rightarrow D^{(*)}$ 
at Zero Recoil}

\begin{figure}[!ht]
\centerline{{\epsfxsize=3.0in \epsfbox{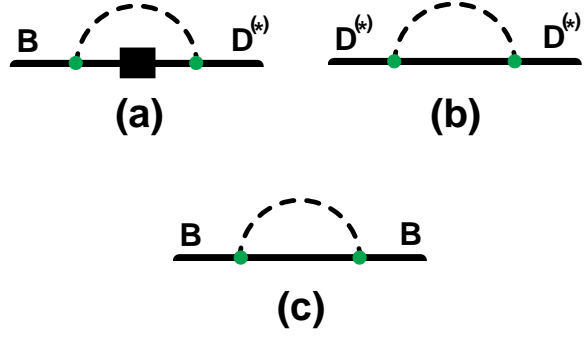}}} 
\vskip 0.15in
\noindent
\caption{\it 
One-loop diagrams that contribute
to the matrix elements for 
$B\rightarrow D^{(*)}$.
The solid lines denote mesons containing a heavy quark, while the dashed lines
denote light mesons.  The solid square corresponds to an insertion of the 
weak interaction vertex, while the small solid circles denote a strong
interaction vertex $\propto g_\pi$.
Diagram (a) is a vertex correction, while diagrams (b) and (c) 
correspond to wavefunction renormalization.
}
\label{fig:IWloops}
\vskip .2in
\end{figure}

The matrix elements for the semileptonic decays 
$B\rightarrow D e\nu$ and $B\rightarrow D^* e\nu$ at zero-recoil
are normalized to unity in the heavy quark
limit~\cite{IW}.
Further, there are no corrections at zero-recoil of the form
$1/m_{c,b}$ by Luke's theorem~\cite{Mikes}, and the leading
corrections in the heavy quark expansion enter at $1/m_c^2$.
In addition to contributions from perturbative insertions
of local operators in the heavy
quark Lagrange density, there are also contributions 
that receive infrared enhancements
from long-distance strong-interactions processes. The largest
contribution of this type is from the one-loop diagrams
shown in fig.~\ref{fig:IWloops}
in which the $D^*-D$ mass-splitting is explicitly retained in the 
$D^{(*)}$ propagator~\cite{RW93}.
Recently, quenched lattice QCD 
simulations of the $B\rightarrow D^{(*)}$ 
form factors have been performed~\cite{Fermi99,UKQCD01} 
but the light-quark mass dependence of the results has yet to be ascertained..

The hadronic matrix elements for $B\rightarrow D^{(*)}e\nu$ are
\begin{eqnarray}
\langle D (v^\prime)|\overline{c}\gamma^\mu b|B(v)\rangle
& = & \sqrt{m_B m_D}\ \left[\ 
h_+ (w)(v+v^\prime)^\mu\ +\ h_-(w) (v-v^\prime)^\mu
\ \right]
\nonumber\\
\langle D^* (v^\prime, \epsilon)|\overline{c}\gamma^\mu \gamma_5b|B(v)\rangle
& = & \sqrt{m_B m_{D^*}}\  \left[\ 
-i h_{A_1}(w) (w+1) \epsilon^{* \mu}
+i h_{A_2}(w) (v\cdot\epsilon^*) v^\mu
\right.\nonumber\\ & & \left.\qquad\qquad\qquad
+i h_{A_3}(w) (v\cdot\epsilon^*) v^{\prime \mu}
\ \right]
\nonumber\\
\langle D^* (v^\prime, \epsilon)|\overline{c}\gamma^\mu b|B(v)\rangle
& = & \sqrt{m_B m_{D^*}}\  h_V(w)\varepsilon^{\mu\nu\alpha\beta}
\epsilon_\nu^*v_\alpha^\prime v_\beta
\ \ \ ,
\label{eq:BDmat}
\end{eqnarray}
where $v_\mu$ is the four-velocity of the initial-state $B$-meson,
$v_\mu^\prime$ is the four-velocity of the final-state $D^{(*)}$-meson,
$w=v\cdot v^\prime$ and $h_\pm$, $h_V$ and $h_{A_{1,2,3}}$
are the six independent form factors that contribute.
In the heavy quark limit $h_- (w)=h_{A_2}(w) =0$ 
and $h_+(w)= h_V(w)=h_{A_{1,3}}(w) =\xi(w)$ where $\xi(w)$ is the universal
function for the $B^{(*)}\rightarrow D^{(*)}$ decays
known as the Isgur-Wise function.
The matrix element is reproduced in the heavy meson chiral lagrangian
by an operator of the form
\begin{eqnarray}
\overline{c}\gamma^\mu(1-\gamma_5) b
&\rightarrow &
-\xi(w) \ {\rm Tr}\left[\ \overline{H}^{(c)}_{v^\prime}
\gamma^\mu (1-\gamma_5)
H^{(b)}_v\ \right]
\ \ \ ,
\label{eq:IWeff}
\end{eqnarray}
where $\xi(w)$ is the Isgur-Wise function, and we have retained the 
flavor labels on the $H$ fields.

The lowest order  heavy-quark-spin-symmetry breaking operator 
in the heavy quark chiral Lagrangian is 
dimension three due to the chromomagnetic moment interactions of
the charm quark with the light degrees of freedom, $\sim
\overline{c}\ \sigma^{\mu\nu} G^{\mu\nu} c/m_c$
(we neglect the contribution from the b-quark chromomagnetic interactions,
i.e. $m_b\gg m_c$).
This interaction gives rise to a mass-splitting between the $D^{*}$
and $D$ at order $1/m_c$ in the heavy meson chiral Lagrangian, which
after performing the Dirac trace, is
\begin{eqnarray}
\delta {\cal L}^{(c)} & = & \Delta^{(c)} D^{*\dagger}_\alpha
D^{*\alpha}
\ \ \ .
\end{eqnarray}
As this is the lowest dimension operator that breaks the heavy quark
symmetries, it provides the dominant long-distance contribution
to the $1/m_c^2$ corrections to the matrix elements in
eq.~(\ref{eq:BDmat})~\cite{RW93}.
Randall and Wise~\cite{RW93} found a contribution of the form 
$\left(\Delta^{(c)}\right)^2\log\left( m_\pi^2/\Lambda^2\right)+...$ 
at $w=1$
from the pion loop-diagrams shown in  fig.~\ref{fig:IWloops}.
In order to exactly compensate 
the renormalization scale dependence, $\Lambda$,
the $1/m_c^2$ operators that are independent of $m_q$ will combine
together to give a $m_q$-independent contribution 
$X_+ (\Lambda)$ at $w=1$ for the $B\rightarrow D$ decay and 
$X_A (\Lambda)$ at $w=1$  for the $B\rightarrow D^*$ decay.
The form factors at zero-recoil are modified at 
${\cal O}\left(1/m_c^2\right)$
by the one-loop diagrams 
in fig.~\ref{fig:IWloops} to become
\begin{eqnarray}
h_{+}^{(B_q) PQ}(1) & = & 1 \ +\  X_+(\Lambda)\ +\ 
{g_\pi^2 \over 16\pi^2 f^2}\ \left[\  
3 F_{jq} \ -\  {1\over 3} F_{qq} 
\ -\  {1\over 3} \left(m_{qq}^2-m_{jj}^2\right) R_{qq}
\ \right]
\nonumber\\
h_{+}^{(B_j) PQ}(1) & = & 1 \ +\  X_+(\Lambda)\ +\ 
{g_\pi^2 \over 6\pi^2 f^2}\ 
F_{jj}
\nonumber\\
h_{A_1}^{(B_q) PQ}(1) & = & 1 \ +\  X_A(\Lambda)\ +\ 
{g_\pi^2 \over 48\pi^2 f^2}\ \left[\ 
3 \overline{F}_{jq} \ -\  {1\over 3} \overline{F}_{qq} 
\ -\  {1\over 3} \left(m_{qq}^2-m_{jj}^2\right) \overline{R}_{qq}
\ \right]
\nonumber\\
h_{A_1}^{(B_j) PQ}(1) & = & 1 \ +\  X_A(\Lambda)\ +\ 
{g_\pi^2 \over 18\pi^2 f^2}\
\overline{F}_{jj}
\ \ \ ,
\label{eq:RWcorr}
\end{eqnarray}
where we have used the shorthand
$F_{ab}=F(m_{ab},\Delta^{(c)},\Lambda)$,
$\overline{F}_{ab}=F(m_{ab},-\Delta^{(c)},\Lambda)$,
$R_{ab}=R(m_{ab},\Delta^{(c)},\Lambda)$ and 
$\overline{R}_{ab}=R(m_{ab},-\Delta^{(c)},\Lambda)$.
The expressions in eq.~(\ref{eq:RWcorr}) are true for $q~=~u,d,s$.
The loop functions $F$ and $R$ are
\begin{eqnarray}
F(m,\delta,\Lambda) & = & \delta^2\log\left({m^2\over\Lambda^2}\right)
\ - \ 2(2m^2-\delta^2)
\ -\ \delta 
\sqrt{\delta^2-m^2}\log\left({\delta-\sqrt{\delta^2-m^2+i\epsilon}\over
\delta+\sqrt{\delta^2-m^2+i\epsilon}}\right)
\nonumber\\
& & 
\ +\ {2 m^2\over\delta}
\left(\pi m - 
\sqrt{\delta^2-m^2}\log\left({\delta-\sqrt{\delta^2-m^2+i\epsilon}\over
\delta+\sqrt{\delta^2-m^2+i\epsilon}}\right)\right)
\nonumber\\
&\rightarrow &
\delta^2 \log\left({m^2\over\Lambda^2}\right) 
\ +\ 
{\cal O}\left(\delta^3\right)
\nonumber\\
R(m,\delta,\Lambda) & = & {3\pi m\over\delta} \ -\ 6
\ -\ {3(\delta^2-2 m^2)\over 2\delta\sqrt{\delta^2-m^2}}
\log\left({\delta-\sqrt{\delta^2-m^2+i\epsilon}\over
\delta+\sqrt{\delta^2-m^2+i\epsilon}}\right)
\nonumber\\
&\rightarrow & {\delta^2\over m^2}\ +\ {\cal O}\left(\delta^3\right)
\ \ \ .
\label{eq:RWloopfun}
\end{eqnarray}
It is interesting to note that in the $\Delta^{(c)}\ll m_{qq}$ limit,
there is a contribution to the $h_{i}^{(B_q) PQ}(1)$ of the form
$(m_{qq}^2-m_{jj}^2)/m_{qq}^2$.
This is analogous to the enhanced chiral logarithms found by Sharpe~\cite{S97}
that contribute to the light meson masses and decay constants.
Once lattice simulations have determined $X_+(\Lambda)$ and 
$X_A(\Lambda)$, they can be inserted
into the QCD expressions for $h_+(1)$ and $h_{A_1}(1)$
for $B_{u,d,s}$
to obtain the zero-recoil matrix elements in QCD,
\begin{eqnarray}
h_+^{(B_u)} (1) & = &  1 \ +\  
X_+(\Lambda)\ +\ {g_\pi^2\over 16\pi^2 f^2} 
\left[\ 
{3\over 2} F_\pi\ +\ F_K\ +\ {1\over 6} F_\eta\ 
\right]
\nonumber\\
h_+^{(B_s)} (1) & = & 1\ +\ 
X_+(\Lambda)\ +\ {g_\pi^2\over 16\pi^2 f^2} 
\left[\ 
2F_K\ +\ {2\over 3} F_\eta\ 
\right]
\nonumber\\
h_{A_1}^{(B_u)} (1) & = &  1 \ +\  
X_A(\Lambda)\ +\ {g_\pi^2\over 48\pi^2 f^2} 
\left[\ 
{3\over 2} \overline{F}_\pi\ +\ \overline{F}_K\ 
+\ {1\over 6} \overline{F}_\eta\ 
\right]
\nonumber\\
h_{A_1}^{(B_s)} (1) & = & 1\ +\ 
X_A(\Lambda)\ +\ {g_\pi^2\over 48\pi^2 f^2} 
\left[\ 
2\overline{F}_K\ +\ {2\over 3} \overline{F}_\eta\ 
\right]
\ \ \ ,
\label{eq:RWQCD}
\end{eqnarray}
where we have used $F_Y=F(m_Y,\Delta^{(c)},\Lambda)$
and $\overline{F}_Y=F(m_Y,-\Delta^{(c)},\Lambda)$.
Keeping only the contribution from pion-loops 
and counterterms we recover the 
results of Ref.~\cite{RW93}.

\section{Chiral Corrections to the $B\rightarrow D$ Isgur-Wise Function}

While the Isgur-Wise function $\xi (w)$ in eq.~(\ref{eq:IWeff})
is normalized to unity  at zero recoil, $\xi (1)=1$, away from zero
recoil its value depends upon the light quark masses.
In QCD the leading non-analytic contributions  to 
the Isgur-Wise function for $B_{u,d}$ decays,
$\xi_{B_{u,d}} (w)$, 
and the Isgur-Wise function for $B_{s}$ decays, $\xi_{B_s} (w)$,
of the form $m_q\log m_q$ have been computed~\cite{JS92a,G92}.
In addition, the contributions of this form arising in quenched QCD (QQCD)
have also been computed~\cite{Booth95}.
In the heavy quark limit, matrix elements 
of the operator $\overline{c}\gamma^\mu(1-\gamma_5)b$
between $B$ and $D^{(*)}$ states are reproduced,
up to ${\cal O}(m_q)$, by
\begin{eqnarray}
 \overline{c}\gamma^\mu(1-\gamma_5)b  \rightarrow & & 
-\xi (w)\ 
{\rm Tr}\left[\ \overline{H}^{(c)}_{v^\prime} 
\gamma^\mu (1-\gamma_5) H^{(b)}_v
\right]
\ -\ \eta_1 (w, \Lambda) 
{\rm Tr}\left[\ \overline{H}^{(c)}_{v^\prime} 
\gamma^\mu (1-\gamma_5) H^{(b)}_v
{\cal M}_+ \ \right]
\nonumber\\ & & 
 \ -\ 
\eta_2 (w, \Lambda) \ 
{\rm str}\left[\ {\cal M}_+\ \right]
{\rm Tr}\left[\ \overline{H}^{(c)}_{v^\prime} 
\gamma^\mu (1-\gamma_5) H^{(b)}_v\ \right]
\ \ \ ,
\label{eq:IWmq}
\end{eqnarray}
where we use notation similar to Ref.~\cite{G92} for the higher order
counterterms, $\eta_{1,2}$, and 
it is understood that flavor indices are super-traced over.
In addition heavy quark symmetry requires that
$\eta_{1,2}(1, \Lambda)=0$
The one-loop diagrams shown in fig.~\ref{fig:IWloops} and the contributions
from eq.~(\ref{eq:IWmq}) give Isgur-Wise functions in QCD at 
${\cal O}\left(m_q\right)$
\begin{eqnarray}
\xi_{B_{u,d}} (w) & = & \xi (w)\ \left[\ 
1 + {g_\pi^2\over 8\pi^2 f^2}\left[\ r(w)-1\ \right]
\left[\ {3\over 2} m_\pi^2\log {m_\pi^2\over \Lambda^2}
+ m_K^2\log {m_K^2\over\Lambda^2}
+{1\over 6} m_\eta^2\log{m_\eta^2\over\Lambda^2}\ 
\right]\ \right]
\nonumber\\
& & +\ \eta_1^{(R)} (w, \Lambda)\  \overline{m} 
\ +\  \eta_2^{(R)}(w, \Lambda) \left(\ 2\overline{m}\ +\ m_s\ \right)
\nonumber\\
\xi_{B_s} (w) & = & \xi (w)\ \left[\ 
1 + {g_\pi^2\over 8\pi^2 f^2}\left[\ r(w)-1\ \right]
\left[\ 2 m_K^2\log {m_K^2\over\Lambda^2}
+{2\over 3}m_\eta^2\log{m_\eta^2\over\Lambda^2}\ 
\right]\ \right]
\nonumber\\
& & +\ \eta_1^{(R)} (w, \Lambda) \ m_s \ +\  
\eta_2^{(R)}(w, \Lambda) \left(\ 2\overline{m}\ +\ m_s\ \right)
\ \ \ \ ,
\label{eq:IWQCD}
\end{eqnarray}
where the $w$-dependent function is~\cite{JS92a,G92}
\begin{eqnarray}
r(w) & = & {1\over\sqrt{w^2-1}}\ \log\left[ w\ +\ \sqrt{w^2-1}\right]
\ \ \ .
\label{eq:rfun}
\end{eqnarray}
$\eta_{1,2}^{(R)}$ are renormalized constants with
$\eta_{1,2}^{(R)}(1, \Lambda)=0$.

In PQ$\chi$PT, the Isgur-Wise functions 
for mesons composed of a b-quark and an anti-valence quark,
$b\overline{u}$,  $b\overline{d}$, or  $b\overline{s}$,
at one-loop order and in the limit of isospin symmetry, are
\begin{eqnarray}
\xi_{B_q}^{PQ} (w) & = & \xi (w)\ \left[
\ 1 + {g_\pi^2\over 8\pi^2 f^2}\left[\ r(w)-1\ \right]
\left[
3 m_{jq}^2\log {m_{jq}^2\over\Lambda^2}
-{1\over 3} (2 m_{qq}^2-m_{jj}^2)\log {m_{qq}^2\over\Lambda^2}
\right]\ \right]
\nonumber\\
& & +\ \eta_1^{(R)} (w, \Lambda)\  m_q + 3\  \eta_2^{(R)} (w, \Lambda)\  m_j 
\ \ \ ,
\label{eq:IWPQQCD}
\end{eqnarray}
where $q=u,d,s$.
The Isgur-Wise function for mesons comprised of a b-quark and an anti-sea
quark, $j$, is
\begin{eqnarray}
\xi_{B_j}^{PQ} (w) & = & \xi (w)\left[
1 + {g_\pi^2\over 3\pi^2 f^2}\left[\ r(w)-1\ \right]
m_{jj}^2 \log {m_{jj}^2\over\Lambda^2}
\ \right]
+\left(\eta_1^{(R)}(w,\Lambda)+3\eta_2^{(R)}(w,\Lambda)
\right)m_j
\ .
\label{eq:IWbsea}
\end{eqnarray}
As required, the expression in eq.~(\ref{eq:IWQCD}) reduces to the expression
in eq.~(\ref{eq:IWbsea}) when $m_\pi=m_K=m_\eta=m_{jj}$ and $m_{u,d,s}=m_j$.
By determining $\xi_{B_q}^{PQ} (w)$ and 
$\xi_{B_j}^{PQ} (w)$ in lattice simulations
and fitting the quantities $\xi(w)$ and  $\eta_{1,2}(w,\Lambda)$, 
one can then use the 
QCD expressions for $\xi_{B_{u,d,s}} (w)$ in eq.~(\ref{eq:IWQCD})
to perform the extrapolation in $m_q$.


\section{Radiative Transitions: $D^*\rightarrow D\gamma$}

The widths of the radiative decays $D^*\rightarrow D\gamma$ are 
relatively well measured. These decays are quite interesting as they have
allowed for a determination of the axial coupling constant $g_\pi$, in
combination with the leading order magnetic transition
moment counterterm~\cite{radmany,Stewart98}.
The value for $g_\pi$ that has been extracted from these decays is
significantly smaller~\cite{radmany,Stewart98} 
than the value expected in the naive quark model.

\begin{figure}[!ht]
\centerline{{\epsfxsize=1.5in \epsfbox{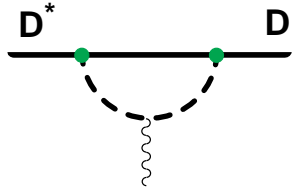}}} 
\vskip 0.15in
\noindent
\caption{\it 
One-loop diagrams that contribute
to the radiative transitions $D^*\rightarrow D\gamma$.
The solid lines denote mesons containing a heavy quark, 
the dashed lines
denote light mesons and 
the small solid-circle denotes a strong 
interaction $\propto g_\pi$.
}
\label{fig:magloops}
\vskip .2in
\end{figure}

The width for radiative decays is given in terms of a total radiative matrix
element $\mu_a$,
\begin{eqnarray}
\Gamma (D_a^*\rightarrow D_a\gamma) & = & 
{\alpha\over 3}\ |\mu_a|^2 \ |{\bf k_\gamma}|^3
\ \ \ ,
\label{eq:magwidth}
\end{eqnarray}
where the subscript ``$a$'' denotes the flavor of the $D^{(*)}$ meson.
At leading order there are contributions that are independent of the light
quark masses arising from the coupling of the photon to the light degrees of
freedom, and to the heavy quark.
At next to leading order there is a contribution of the form $\sqrt{m_q}$
resulting from the one-loop diagrams shown in fig.~\ref{fig:magloops}.
In the heavy quark limit, the transition moments in QCD are 
\begin{eqnarray}
\mu_u & = & {2\over 3 m_c}\ +\ {2\over 3}\beta\ -\ {g_\pi^2 m_K\over 4\pi f^2}
\ -\ {g_\pi^2 m_\pi\over 4\pi f^2}
\nonumber\\
\mu_d & = & {2\over 3 m_c}\ -\ {1\over 3}\beta\  
+\ {g_\pi^2 m_\pi\over 4\pi f^2}
\nonumber\\
\mu_s & = & {2\over 3 m_c}\ -\ {1\over 3}\beta\  
+\ {g_\pi^2 m_K\over 4\pi f^2}
\ \ \ ,
\label{eq:magmoms}
\end{eqnarray}
where $\beta$ is the $m_q$-independent
counterterm contribution from the light degrees of
freedom.

In PQQCD the calculations are analogous to those of QCD, and one finds that
\begin{eqnarray}
\mu_u^{PQ} & = & 
{2\over 3 m_c}\ +\ {2\over 3}\beta\ -\ {g_\pi^2 m_{ju}\over 2\pi f^2}
\nonumber\\
\mu_d^{PQ} & = & {2\over 3 m_c}\ -\ {1\over 3}\beta\  
+\ {g_\pi^2 m_{ju}\over 4\pi f^2}
\nonumber\\
\mu_s^{PQ} & = & {2\over 3 m_c}\ -\ {1\over 3}\beta\  
+\ {g_\pi^2 m_{js}\over 4\pi f^2}
\nonumber\\
\mu_j^{PQ} & = & {2\over 3 m_c}\ +\ {2\over 3}\beta\  
-\ {g_\pi^2 m_{jj}\over 2\pi f^2}
\ \ \ ,
\label{eq:magmomsPQ}
\end{eqnarray}
where the coefficient $\beta$ is the same in QCD and PQQCD.
A fit to  lattice calculations of these matrix elements will provide
a determination of $\beta$, and consequently an estimate of the radiative
matrix elements for the physical values of $m_q$.

\section{Matrix Elements of Isovector Twist-2 Operators}

Recently, it has been realized that the chiral corrections to the matrix
elements of isovector twist-2 operators in the nucleon and octet-baryons
can be computed in 
chiral perturbation theory~\cite{ChJi,ArSa,ChSa,Chenmorea,Chenmoreb}.
Such corrections systematically incorporate the long-distance strong
interaction contributions to the moments of the parton distributions.
While it is not possible to perform deep-inelastic-scattering (DIS) 
from hadrons containing heavy quarks, it is possible to consider the parton
distributions of heavy hadrons from a theoretical standpoint, and further it
is likely that lattice studies of these distributions will be significantly
easier than analogous studies in nucleons.

In QCD the twist-2 isovector operators have the form
\begin{eqnarray}
{\cal O}^{(n), a}_{\mu_1\mu_2\ ... \mu_n}
& = & 
{1\over n!}\ 
\overline{q}\ \lambda^a\ \gamma_{ \{\mu_1  } 
\left(i \stackrel{\leftrightarrow}{D}_{\mu_2}\right)\ 
... 
\left(i \stackrel{\leftrightarrow}{D}_{ \mu_n\} }\right)\ q
\ -\ {\rm traces}
\ \ \ ,
\label{eq:twistop}
\end{eqnarray}
where the $\{ ... \}$ denotes symmetrization on all Lorentz indices,
and $\lambda^a$ are $SU(3)$ Gell-Mann matrices.
In PQQCD the operator has an analogous form,
\begin{eqnarray}
{\cal O}^{(n), a}_{\mu_1\mu_2\ ... \mu_n}
& = & 
{1\over n!}\ 
\overline{Q}\ \overline{\lambda}^a\ \gamma_{ \{\mu_1  } 
\left(i \stackrel{\leftrightarrow}{D}_{\mu_2}\right)\ 
... 
\left(i \stackrel{\leftrightarrow}{D}_{ \mu_n\} }\right)\ Q
\ -\ {\rm traces}
\ \ \ ,
\label{eq:twistopP}
\end{eqnarray}
where $\overline{\lambda}^a$ is a super Gell-Mann matrix.
As an example, the $\overline{\lambda}^3$ matrix has entries  
${\rm diag}(1,-1,0,1,-1,0,1,-1,0)$.
Forward matrix elements of
${\cal O}^{(n), 3}_{\mu_1\mu_2\ ... \mu_n}$
between heavy meson states are reproduced up to ${\cal O}(m_q)$ 
by forward matrix elements of 
\begin{eqnarray}
{\cal O}^{(n), 3}_{\mu_1\mu_2\ ... \mu_n} \  \rightarrow\  
& & -{1\over 2} v_{\mu_1}v_{\mu_2}...v_{\mu_n}
\left[\ 
t^{(n)} 
{\rm Tr}\left[\ \overline{H}_v H_v \overline{\lambda}^3_{\xi+}\ \right]
\ +\ 
s_1^{(n)}(\Lambda)
{\rm Tr}\left[\ \overline{H}_v H_v 
{\cal M}_+ \overline{\lambda}^3_{\xi+}\ \right]
\right.\nonumber\\ & & \left.
\ +\ 
s_2^{(n)}(\Lambda)
{\rm Tr}\left[\ \overline{H}_v H_v \overline{\lambda}^3_{\xi+}  {\cal M}_+ 
\ \right]
\ +\ 
s_3^{(n)}(\Lambda)
{\rm Tr}\left[\ \overline{H}_v H_v \overline{\lambda}^3_{\xi+} \ \right]
{\rm str}\left[\   {\cal M}_+\ \right]\ 
\right]
\ \ \ ,
\label{eq:TTops}
\end{eqnarray}
where 
$\overline{\lambda}^3_{\xi+} = {1\over 2}
\left(\xi^\dagger\overline{\lambda}^3\xi + \xi
  \overline{\lambda}^3\xi^\dagger\right)$,
and it is understood that flavor indices are super-traced over.
\begin{figure}[!ht]
\centerline{{\epsfxsize=2.5in \epsfbox{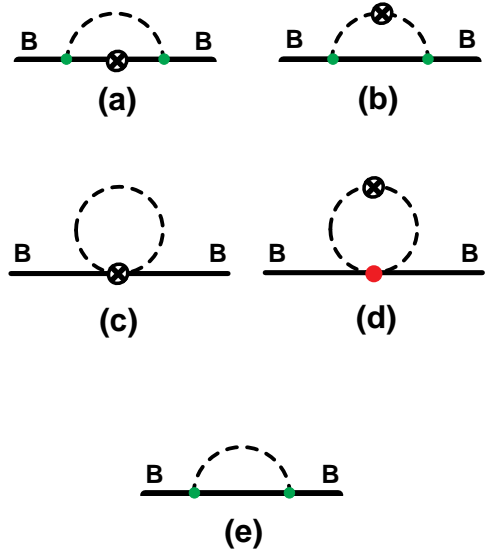}}} 
\vskip 0.15in
\noindent
\caption{\it 
One-loop diagrams that contribute
to the matrix elements of the isovector 
twist-2 operators in mesons containing
a heavy quark.
The solid lines denote mesons containing a heavy quark, 
while the dashed lines denote light mesons. 
The crossed-circle denotes an insertion of 
${\cal O}^{(n), a}_{\mu_1\mu_2\ ... \mu_n} $
while the small solid-circle denotes a strong interaction $\propto g_\pi$.
}
\label{fig:twistloops}
\vskip .2in
\end{figure}
At ${\cal O}\left(m_q\right)$ in the chiral expansion 
the forward matrix elements receive
contributions from the operators in eq.~(\ref{eq:TTops}) and the one-loop
diagrams shown in fig.~\ref{fig:twistloops}, 
\begin{eqnarray}
\langle B_u|{\cal O}^{(n), 3}_{\mu_1\mu_2\ ... \mu_n}|B_u\rangle^{PQ}
 = \
 v_{\mu_1}v_{\mu_2}...v_{\mu_n}
& & \left[\ 
t^{(n)}\left(\ 1 - {3(1+3 g_\pi^2)\over16\pi^2 f^2}
  m_{ju}^2\log{m_{ju}^2\over\Lambda^2}\ \right)
\right.\nonumber\\ & & \left.
\ +\ \left(\ s_1^{(R)(n)}(\Lambda)
\ +\ s_2^{(R)(n)}(\Lambda)\right)\ \overline{m}
\ +\ 3 s_3^{(R)(n)}(\Lambda)\ m_j\ \right]
\nonumber\\
\langle B_j|{\cal O}^{(n), 3}_{\mu_1\mu_2\ ... \mu_n}|B_j\rangle^{PQ}
 = \
 v_{\mu_1}v_{\mu_2}...v_{\mu_n}
& & \left[\ 
t^{(n)}\left(\ 1 - {3(1+3 g_\pi^2)\over16\pi^2 f^2}
  m_{jj}^2\log{m_{jj}^2\over\Lambda^2}\ \right)
\right.\nonumber\\ & & \left.
\ +\ \left(\ s_1^{(R)(n)}(\Lambda)\ +\ s_2^{(R)(n)}(\Lambda)
\ +\ 3 s_3^{(R)(n)}(\Lambda)\ \right)\ m_j\ \right]
\ \ \ ,
\label{eq:Btwist}
\end{eqnarray}
for $n={\rm odd}$ and $n>1$.
In the loop calculations we have only retained the non-analytic contributions,
and the $s_i^{(R)(n)}$ are renormalized counterterms.
For $n=1$ the matrix element is absolutely normalized since
${\cal O}^{(1), 3}_{\mu_1}$ is the isovector charge operator.
Matrix elements in the other meson states can be found
straightforwardly from those in eq.~(\ref{eq:Btwist}).
From PQQCD lattice simulations one envisages that the constants
$t^{(N)}$, $s_3^{(R)(n)}(\Lambda)$ and the combination 
$s_1^{(R)(n)}(\Lambda)\ +\ s_2^{(R)(n)}(\Lambda)$
can be individually determined and then inserted into the analogous 
expressions in QCD to determine the moments of combinations of
parton distributions in heavy mesons.
In QCD I find the matrix element between $B_u$ states to be
\begin{eqnarray}
\langle B_u|{\cal O}^{(n), 3}_{\mu_1\mu_2\ ... \mu_n}|B_u\rangle
& &\ =\   v_{\mu_1}v_{\mu_2}...v_{\mu_n}
\left[\ 
t^{(n)}\left(\ 1 - {(1+3 g_\pi^2)\over16\pi^2 f^2}\left(\ 
  2 m_{\pi}^2\log{m_{\pi}^2\over\Lambda^2}\ +\ 
m_K^2\log{m_{K}^2\over\Lambda^2}
\ \right)\ \right)
\right.\nonumber\\ & & \left.
+\left(\ s_1^{(R)(n)}(\Lambda)+s_2^{(R)(n)}(\Lambda)
+ 2 s_3^{(R)(n)}(\Lambda)\right)\ \overline{m}
\ +\ s_3^{(R)(n)}(\Lambda)\ m_s\ \right]
\ .
\label{eq:BtwistQCD}
\end{eqnarray}

\section{Conclusions}

In this work I have computed several observables in mesons containing heavy
quarks at one-loop level in partially quenched heavy quark
chiral perturbation theory and
in ordinary heavy quark chiral perturbation theory.
In order to determine these observables from numerical lattice QCD simulations,
an extrapolation in the quark masses from those that can be simulated 
on the lattice down to their physical values is required.
The recent progress in understanding partially quenched QCD has provided 
an efficient pathway to perform such extrapolations.
The counterterms that contribute to partially quenched observables are 
the same as those that contribute to the analogous QCD observables, and hence
partially quenched simulations along with the light-quark mass dependence 
about the chiral limit are sufficient to recover the QCD prediction to some
order in the chiral  expansion.

\bigskip\bigskip

MJS would also like to thank the Aspen Institute for Physics for providing the
wonderful environment in which this work was performed.
I would also like to thank the organizers of the heavy flavors workshop that
was held in Aspen during August and September of 2001.
This work is supported in
part by the U.S. Dept. of Energy under Grants No.  DE-FG03-97ER4014.

\end{document}